\documentclass[%
 reprint,
 amsmath,amssymb,
 aps,
]{revtex4-2}

\usepackage{graphicx}
\usepackage{dcolumn}
\usepackage{bm}
\usepackage{xcolor}

\usepackage[colorlinks=true,
            linkcolor=black,   
            citecolor=blue,    
            urlcolor=blue]     
            {hyperref}

\begin{document}

\preprint{APS/123-QED}

\title{Stability of current-carrying states in hard-core bosons with long-range hopping on a square lattice
 }

\author{Yoshihiro Yabuuchi$^{1,2}$}
\email[]{sf23347z@st.omu.ac.jp}

\author{Ippei Danshita$^{2}$}%
 \email[]{danshita@phys.kindai.ac.jp}

\affiliation{$^{1}$Department of Physics, Osaka Metropolitan University, Sugimoto, Osaka 558-8585, Japan \\ $^{2}$Department of Physics, Kindai University, Higashi-Osaka, Osaka 577-8502, Japan}

\date{\today}

\begin{abstract}
We investigate the stability of current-carrying states with quasi-momentum $K$ in the Bose-condensed phase of the hard-core Bose-Hubbard model on a square lattice, where particles transfer between two sites separated by distance $r$ with hopping amplitude decaying algebraically with $r$ as $\propto r^{-\alpha}$. 
Using a mean-field theory, we analyze the excitation spectrum and determine the critical quasi-momenta associated with Landau and dynamical instabilities. 
We find that the long-range hopping suppresses the critical quasi-momenta and makes them vanish at $\alpha=3$.
Near $\alpha=3$, we show that the critical quasi-momentum $K_{\mathrm{c}}$ for the dynamical instability exhibits the scaling behavior $K_\mathrm{c} \propto \Delta^{1+\Delta}$ with $\Delta=\alpha-3$, where the scaling exponent explicitly depends on $\Delta$, as a consequence of the long-range nature of the hopping. 

\end{abstract}

\maketitle


\section{\label{introduction} Introduction}

Superfluidity, which is the phenomenon of frictionless flow, has been first found in liquid helium systems~\cite{Kapitza1938,Allen1938} and been a central theme in quantum many-body physics for many decades~\cite{Svistunov,Pitaevskii}.
The realization of Bose-Einstein condensates (BEC) of neutral atomic Bose gases in 1995~\cite{Davis1995,Anderson1995} have pioneered a new era of the research on superfluidity in the platform of atomic, molecular, and optical (AMO) physics~\cite{Pitaevskii}. Various superfluid properties have been indeed observed in atomic BEC systems, such as quantized vortices~\cite{Madison2000,Aboshaeer2001}, the superfluid fraction~\cite{Chauveau2023}, persistent currents in ring traps~\cite{Ramanathan2011,Wright2013}, and superfluid critical velocities~\cite{Raman1999,Desbuquois2012}. Furthermore, the usage of optical lattices has enriched the studies of superfluidity~\cite{Cataliotti2001,Burger2001,Greiner2002,Fallani2004,Desarlo2005,Fertig2005,Desarlo2005,Mun2007,McKay2008,Haller2010,Gadway2011,Tanzi2016}, stimulating theoretical investigation of that in lattice-boson systems~\cite{Smerzi2002,Wu_2003,Altman2005,Polkovnikov2005,Konabe_2006,Iigaya2006,Snoek2007,Danshita2010,Yamamoto_2011,Saito2012}, which is in general described by Bose-Hubbard-type models~\cite{Jaksch1998,Fisher1989}. In particular, the superfluid critical velocities of lattice bosons have been quantitatively measured by means of moving optical lattices~\cite{Fallani2004,Desarlo2005,Mun2007}.

Although the superfluidity in lattice-boson systems has been extensively studied in previous literature, the influence of modifications in the spatial form of the hopping remains to be an intriguing open question.
Recent progress in the AMO platforms, including Rydberg atoms in optical tweezer arrays~\cite{Sylvain2019,chen2023continuous,Chen2025} and trapped ions~\cite{Richerme2014,Jurcevic2014,kotibhaskar2024programmable}, has allowed for realizing quantum many-body systems described by the spin-1/2 $XY$ model with long-range spin-exchange interactions, in which the coupling strength decays algebraically with distance.
The theoretical mapping between the spin-1/2 $XY$ model and the hard-core Bose-Hubbard model is well known~\cite{Matsubara1956,deSousa2005}, showing that the long-range spin-exchange interaction corresponds to long-range hopping of hard-core bosons (HCB), where the hopping amplitude also decays algebraically with distance.
Such long-range hopping is in stark contrast with the short-range one of the Bose-Hubbard model describing Bose gases in optical lattices, which is dominated by the contribution between the nearest-neighboring sites~\cite{Jaksch1998,Greiner2002}.
In this sense, the technological progress in the AMO platforms have opened up new possibilities for exploring emergent bosonic phenomena arising from the long-range hopping.
Previous studies of this model have revealed that the long-range interaction fundamentally alters the shape of the low-energy excitation spectrum~\cite{Peter2012,Diessel2023,Chen2025}.
These results suggest that the long-range hopping can significantly modify the superfluidity of HCB on a lattice because the excitation spectrum is in general closely related to the linear stability of current-carrying states of BEC~\cite{Svistunov,Pitaevskii}.

In this work, we present a linear stability analysis of current-carrying states with quasi-momentum $K$ in the BEC phase of the hard-core Bose-Hubbard model on a square lattice, where particles transfer between two sites separated by distance $r$ with hopping amplitude decaying in a power law as $\propto r^{-\alpha}$.
Within the framework of a mean-field theory applied to the equivalent spin-1/2 $XY$ model, we analytically calculate the ground-state phase diagram, which consists of the BEC and Mott-insulator phases, and the excitation spectrum for the former phase in the presence of a finite current. From the excitation spectrum, we determine the critical quasi-momenta associated with Landau instability (LI) and dynamical instability (DI) as functions of the decay exponent $\alpha$.
We find that the critical quasi-momenta decreases with decreasing $\alpha$ from a large value and become zero at $\alpha \leq 3$, meaning that they are zero for the case of Rydberg-atom arrays~\cite{Sylvain2019,chen2023continuous,Chen2025}, i.e., $\alpha=3$. 
Furthermore, we analytically show that the critical quasi-momentum $K_{\text{c}}$ for DI near $\alpha = 3$ obeys the scaling law $K_{\text{c}} \propto \Delta^{1+\Delta}$, where $\Delta = \alpha - 3$.
This unconventional critical behavior, in which the critical exponent explicitly depends on $\Delta$, reflects the nature of the long-range hopping.

The remainder of this paper is organized as follows.
In Sec.~\ref{Model}, we introduce the model Hamiltonian describing HCB with long-range hopping on a square lattice.
In Sec.~\ref{Method}, we formulate our mean-field approach, following Ref.~\cite{Danshita2010}.
In Sec.~\ref{sec:Results}, we first show the ground-state phase diagram taking $\alpha$ and the chemical potential as variables.
We next perform a linear stability analysis to determine the critical quasi-momenta for LI and DI.
Finally, in Sec.~\ref{Summary}, we conclude the paper with summary and outlook.

\section{\label{Model} Model}

We consider a hard-core Bose-Hubbard model on a square lattice,
\begin{eqnarray}
\hat{H} = -\sum_{j<l}J_{j,l}(\alpha)\left( \hat{b}_j^\dagger\hat{b}_l +{\rm h.c.} \right) -\mu\sum_{j}\hat{n}_j,
\label{eq:HCB}
\end{eqnarray}
which possesses the hopping amplitude decaying algebraically with the distance $|{\bf r}_j-{\bf r}_l|$ between two sites $j$ and $l$
\begin{eqnarray}
J_{j,l}(\alpha)=J\left(\frac{a}{|{\bf r}_j-{\bf r}_l|}\right)^\alpha.
\end{eqnarray}
Here, $a$ is the lattice spacing, $\mu$ the chemical potential, and $J$ the hopping amplitude between two nearest-neighboring sites. 
The operators $\hat{b}_{j}$ and $\hat{n}_{j}$ denote the annihilation and number operators for an HCB at site $j$, respectively.
On each site, the hard-core constraint imposes $(\hat{b}_{j}^{\dagger})^{2}=\hat{b}_{j}^{2}=0$, and the operators $\hat b_{j}$ and $\hat{b}_{j}^{\dagger}$ satisfy the anticommutation relation $\{\hat b_{j}, \hat{b}_{j}^{\dagger} \}=1$.
In contrast, operators on different sites commute, i.e., $[\hat{b}_{j}, \hat{b}_{l}^{\dagger} ]=0$ for $j\neq l$~\cite{FRIEDBERG199352}.
Throughout this paper, the index $j<l$ in a sum indicates that the sum runs over all distinct pairs of sites $(j, l)$ on the square lattice, whereas $j\neq l$ denotes a sum over all sites $l$ except $l=j$ (for fixed $j$).
By replacing the HCB operators with the $S=1/2$ spin operators as
\begin{equation}
\hat{S}^{-}_{j} = \hat{b}_{j}, \quad
\hat{S}^{z}_{j} = \hat{n}_{j} - \tfrac{1}{2},
\label{Mapping}
\end{equation}
one can see that this model is equivalent to the $S=1/2$ $XY$ model with long-range spin-exchange interactions,
\begin{equation}
\hat{H} = 
- \sum_{j<l}J_{j,l}
\left( \hat{S}^{+}_{j}\hat{S}^{-}_{l} + \hat{S}^{-}_{j}\hat{S}^{+}_{l} \right)
- h \sum_{j} \hat{S}^{z}_{j},
\label{eq:XY}
\end{equation}
where the longitudinal magnetic field $h$ is equal to $\mu$~\cite{Matsubara1956}.

Quantum many-body systems described by this model have been realized with Rydberg-atom arrays~\cite{Sylvain2019,chen2023continuous,Chen2025} and trapped ions~\cite{Richerme2014,Jurcevic2014,kotibhaskar2024programmable}.
In Rydberg-atom arrays, resonant dipole-dipole interactions between opposite-parity Rydberg states naturally lead to the long-range $XY$ model in Eq.~(\ref{eq:XY}) with $\alpha=3$~\cite{Sylvain2019,chen2023continuous,Chen2025}, while van der Waals interactions arising from off-resonant energy shifts can be neglected when the interatomic distance is sufficiently large~\cite{leseleucdekerouara:tel-02088297}.
In trapped-ion systems, long-range spin-exchange couplings can be engineered with a tunable decay exponent in the range $0<\alpha\le 3$~\cite{Richerme2014,Jurcevic2014,kotibhaskar2024programmable}.
Although $\alpha> 3$ is infeasible in these platforms with currently available techniques except for $\alpha =6$, we also analyze the case of $\alpha> 3$ for theoretical interest. Notice that in the limit $\alpha \to \infty$, only the nearest-neighbor hopping survives and such a situation can be recreated with ultracold Bose gases in optical lattices~\cite{Jaksch1998,Greiner2002}.

Since the commutation relations among the spin operators are a little better known than those of the HCB operators, we formulate our analysis in terms of the $XY$ model with long-range spin-exchange interactions in the followings sections.
 
\section{\label{Method} METHOD}

We use a mean-field (MF) theory in order to qualitatively analyze the ground-state phase diagram and the linear stability of the current-carrying states in the Bose-condensed phase of the model of Eq.~(\ref{eq:HCB}), or equivalently the $XY$ ferromagnetic phase of Eq.~(\ref{eq:XY}).
In Refs.~\cite{Peter2012,Diessel2023}, MF approaches have been applied to the $XY$ model with long-range spin-exchange interactions for analyzing its ground-state properties and low-energy excitations qualitatively.
As we will see in Secs.~\ref{Ground state phase diagram} and~\ref{Excitation spectrum}, our analysis for the cases with no current reproduces the results reported in these earlier works.
We note that for $\alpha \ge 4$ and $T>0$, long-range order is prohibited 
by the Mermin-Wagner theorem extended to systems with long-range interactions~\cite{Bruno2001}.
Instead, this regime is characterized by Berezinskii-Kosterlitz-Thouless physics, with quasi-long-range-ordered and disordered phases~\cite{PhysRevLett.127.156801,PhysRevB.106.014106}. 
In this regime, the MF description is generally invalid, even at a qualitative level.
Hence, our mean-field analyses are qualitatively valid in the regime of $\alpha<4$ at finite but sufficiently low temperatures and that of $\alpha\geq 4$ at zero temperature, where the system is in the long-range-ordered phase.

To formulate the MF description of our model, we follow Ref.~\cite{Danshita2010}, 
where the MF approach was applied to systems with short-range spin-exchange 
interactions and a linear stability analysis was performed for current-carrying 
superfluid and supersolid states.
We first introduce the local eigenstates $\left| \uparrow \right\rangle_{j}$ and 
$\left| \downarrow \right\rangle_{j}$ of $\hat{S}_{j}^{z}$, defined by
\begin{align}
\hat{S}_{j}^{z} \left|\uparrow \right\rangle_{j}
= \frac{1}{2} \left| \uparrow \right\rangle_{j}, \qquad
\hat{S}_{j}^{z} \left| \downarrow \right\rangle_{j}
= -\frac{1}{2} \left| \downarrow \right\rangle_{j}.
\end{align}
Here and throughout, we set $\hbar=1$.
Within the MF framework, the wave function of the system is given by the following product state~\cite{PhysRevB.65.104519}:
\begin{equation}
 \left|\Psi_{ \text{MF} } \right\rangle = \bigotimes_{j}\left\{ e^{ -i\frac{ \phi_{j} }{ 2 } } \cos \frac{ \theta_{j} }{ 2 } ~ \left| \uparrow \right\rangle_{ j } + e^{ i\frac{ \phi_{j} }{ 2 } } \sin \frac{ \theta_{j} }{ 2 } ~ \left|\downarrow \right\rangle_{ j } \right\}~,
\label{eq:wave function}
\end{equation}
from which the expectation value of the spin operator at site $j$ is calculated as
\begin{equation}
 \begin{split}
\langle \hat{ \mathbf{S} }_{j} \rangle 
& = \left( \langle \hat{S}_{j}^{x} \rangle,~\langle \hat{S}_{j}^{y} \rangle,~\langle \hat{S}_{j}^{z} \rangle \right) 
\\&= \tfrac{1}{2} \left(  \sin \theta_{j} \cos \phi_{j},~\sin \theta_{j} \sin \phi_{j},~\cos \theta_{j} \right).
 \end{split}
\label{eq:MFspin}
\end{equation}
Here, $\theta_{j}$ and $\phi_{j}$ denote the polar and azimuthal angles of the classical spin at site $j$, respectively. 
This type of MF ansatz has also been employed also in most recent studies of $S=1/2$ quantum spin models, such as the Ising model with mixed fields~\cite{PhysRevA.110.033319} and the $XXZ$ model with long-range interactions~\cite{10.21468/SciPostPhys.17.4.111,kunimi2026magneticfieldcontrolinteractionsalkalineearth}
From Eq.~(\ref{eq:MFspin}), one obtains 
$\langle \hat{S}^{\pm}_{j} \rangle = \tfrac{1}{2}\sin \theta_{j} e^{ \pm i \phi_{j} }$. 
Using these relations, the MF energy 
$\mathcal{H}_{\text{MF}} \equiv \langle \hat{H} \rangle$ is expressed as
\begin{equation}
\begin{split}
\mathcal{H}_{\text{MF}} 
= -\frac{1}{2}\sum_{j<l} J_{j,l} (\alpha)\sin\theta_{j}\sin\theta_{l}\cos\phi_{jl}
   -\frac{h}{2}\sum_{j}\cos\theta_{j},
   \label{MFenergy}
\end{split}
\end{equation}
where $\phi_{jl}\equiv\phi_{j} - \phi_{l}$.
Notice that the number of the bosons $n_j=\langle \hat{n}_j \rangle$ at site $j$ is related to $\theta_j$ as $n_j=(1+\cos\theta_j)/2$.
Replacing the spin operators in the Heisenberg equation of motion with their mean fields of Eq.~(\ref{eq:MFspin}) leads to the following MF equations of motion for $\theta_{j}$ and $\phi_{j}$:
\begin{equation}
\frac{d \theta_{j}}{dt} 
= \sum_{l\neq j}J_{j,l}(\alpha)\sin\theta_{l}\sin\phi_{jl},
\label{motion_theta}
\end{equation}
\begin{equation}
\frac{d \phi_{j}}{dt} 
= \sum_{l\neq j}J_{j,l}(\alpha)
   \frac{\sin\theta_{l}\cos\theta_{j}}{\sin\theta_{j}} \cos\phi_{jl} - h.
\label{motion_phi}
\end{equation}

To investigate the stability of steady states, we express the time-dependent angles as small deviations from their stationary values:
\begin{equation}
    \theta_{j}(t) = \bar{ \theta }_{j} + \delta \theta_{j} e^{ -i \omega t } ~ , ~ 
    \phi_{j}(t) = \bar{ \phi }_{j} + \delta \phi_{j} e^{ -i \omega t }.
    \label{fluctuating motion}
\end{equation}
Here, $\delta\theta_{j}$ and $\delta\phi_{j}$ denote small deviations of $\theta_{j}$ and $\phi_{j}$ from their stationary values $\bar{\theta}_{j}$ and $\bar{\phi}_{j}$, and $\omega$ is the frequency of the normal mode.
The steady-state solution satisfies the following equations : 
\begin{equation}
 \sum_{l\neq j}J_{j,l}(\alpha)\sin \bar{ \theta }_{l}\sin\bar{ \phi }_{jl} = 0 ,
\label{steady_equation_theta}
\end{equation}
\begin{equation}
h = \sum_{l\neq j}J_{j,l}(\alpha)
   \frac{ \sin\bar{ \theta }_{l} \cos\bar{ \theta }_{j} }{ \sin\bar{ \theta }_{j} } \cos\bar{ \phi }_{jl} .
\label{steady_equation_phi}
\end{equation}
Substituting Eq. (\ref{fluctuating motion}) into Eqs.~(\ref{motion_theta}) and (\ref{motion_phi}) and linearizing them with respect to fluctuations, we obtain 

\begin{equation}
\begin{split}
 -i \omega \delta \theta_{j} = \sum_{l\neq j} J_{j,l}(\alpha) \Biggl[ 
 \delta\phi_{jl} \sin \bar{ \theta }_{l} & \cos\bar{ \phi }_{jl} 
 \\ &\!\!\!\!\!\!\!\!\!\! + \delta \theta_{l} \cos\bar{ \theta }_{l} \sin\bar{ \phi }_{jl}
 \Biggr] ,
 \end{split}
\label{fluctuation_equation_theta}
\end{equation}
\begin{equation}
 \begin{split}
\!-i \omega \delta \phi_{j} \!=\! \sum_{l\neq j} \! J_{j,l}(\alpha)  \! &
\Biggl[ \!
  \biggr(
    \delta \theta_{l} \frac{ \cos\bar{ \theta }_{j} \cos\bar{ \theta }_{l} }{ \sin\bar{ \theta }_{j} } \!-\! \delta \theta_{j} \frac{ \sin \bar{ \theta }_{l} }{ \sin^{2} \bar{ \theta }_{j} }
  \biggr) \! \cos\bar{ \phi }_{jl} 
  \\
  &  - \delta \phi_{jl} \frac{ \cos \bar{ \theta }_{j} \sin \bar{ \theta }_{l} \sin \bar{ \phi }_{jl} }{ \sin \bar{ \theta }_{j} }
\Biggr] .
 \end{split}
\label{fluctuation_equation_phi}
\end{equation}

Solving the eigenvalue problem given by Eqs.~(\ref{fluctuation_equation_theta}) and (\ref{fluctuation_equation_phi}), one obtains the frequencies of normal modes as the eigenvalues.
The stability of a steady solution can be judged by $\omega$~\cite{Pitaevskii,Polkovnikov2005,PhysRevA.64.061603}. 
If there are normal modes with the frequencies satisfying $\text{Re}\left[ \omega\right] < 0$ and $\mathrm{Im}\left[ \omega \right] = 0$, the steady solution is energetically unstable. This instability is also called the LI. 
If there are normal modes with the frequencies satisfying $\mathrm{Im}\left[ \omega\right] \neq 0$, the steady state is dynamically unstable.
The emergence of such normal modes with complex frequencies means that they grow exponentially in time, leading to the DI. 
Notice that the LI can destabilize the steady state only when the system is coupled with a kind of thermal bath while the DI can do so even in an isolated system~\cite{Desarlo2005,Konabe_2006,Iigaya2006}.

\section{Results}
\label{sec:Results}
\subsection{\label{Ground state phase diagram} Ground state phase diagram for $\mathbf{K} = \mathbf{0}$}
\label{Ground state phase diagram}

\begin{figure}

\includegraphics[width=1.0\linewidth]{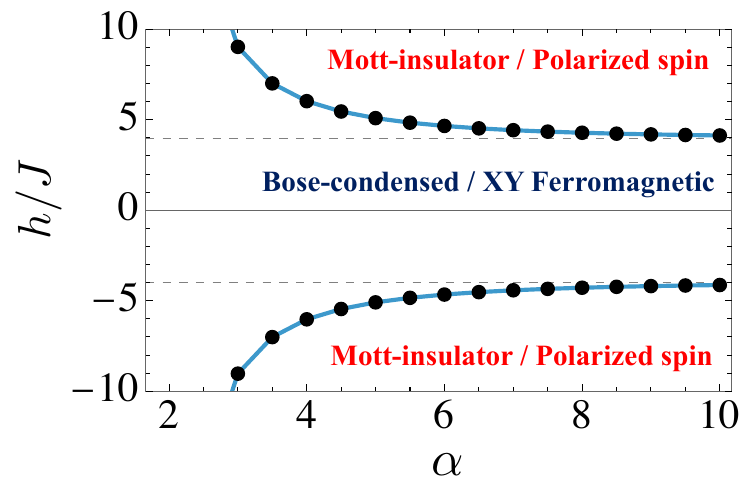}
\caption{Ground-state phase diagram for $\mathbf{K}=\mathbf{0}$: Critical field $\left(h/J\right)_{\text{c}}$ versus the long-range interaction exponent $\alpha$. 
The inner region represents the $XY$ ferromagnetic phase, while outer region corresponds to the polarized phase induced by the strong field $h/J$.
In terms of the HCB, these phases correspond to the Bose-condensed phase and the Mott-insulator phase, respectively. 
The thin dashed lines represent $\left(h/J\right)=\pm 4$
\label{phasediagramK=0}}
\end{figure}

Since we aim to investigate the stability of the current-carrying states in the Bose-condensed phase, we need to locate the parameter region of the Bose-condensed phase in the ground-state phase diagram.
Hence, we first derive the ground-state phase diagram in the $(\alpha,h/J)$-plane, where there exist the Bose-condensed ($XY$ ferromagnetic) phase and the Mott insulator (spin-polarized) phase in terms of the HCB ($S=1/2$ spin).
Without loss of generality we set $\phi_{j} = 0$. 
From Eq.~(\ref{MFenergy}), the MF energy per site of the homogeneous steady state ($\bar{\theta}_j=\bar{\theta}$) is given by
\begin{equation}
    e\left(\bar{\theta}\right)= \frac{\mathcal{H}_{\rm MF}}{M} = -\frac{J}{4}\sin^{2}\bar{\theta}~ \beta_{\alpha}-\frac{h}{2}\cos\bar{\theta},
\end{equation}
with
\begin{equation}
    \beta_{\alpha}=\sum_{l\neq0}\left( \frac{a}{|{\bf r}_{l}|} \right)^{\alpha}.
\end{equation}
where ${\bf r}_{l}=0$ at $l=0$.
The stationary condition $\frac{de}{d\bar{\theta}} = 0$ yields 
$\bar{\theta} = 0,~\pi$, and $\arccos\left(\frac{h}{J\beta_{\alpha}}\right)$. 
The first two solutions correspond to fully polarized states, 
while the last one corresponds to an $XY$-ferromagnetic ordered state with spontaneously breaking of the U$(1)$ symmetry. 
In terms of the HCB, these states correspond to Mott insulating phases with the filling factor $n=1$ and $n=0$,  
and a Bose-condensed phase, respectively. 
Among these candidates, the one that minimizes the MF energy is the ground state for given $h/J$ and $\alpha$, 
and the transition field $\left(h/J\right)_{\text{c}}$ is obtained by equating their energies:
\begin{equation}
    \left( \frac{h}{J}\right)_{\text{c}}=\pm \beta_{\alpha}.
    \label{funcW}
\end{equation}

Figure~\ref{phasediagramK=0} shows the critical field $\left(h/J \right)_{\text{c}}$ as a function of $\alpha$. 
As $\alpha$ decreases, the ordered region expands and $\left(h/J \right)_{\text{c}}$ diverges at $\alpha \rightarrow 2$.
It demonstrates that long-range spin-spin interactions enhance the robustness of the $XY$ ferromagnetic order, which qualitatively agrees with the results of Refs.~\cite{chen2023continuous,Peter2012,deSousa2005}. 
In terms of the HCB, this robustness corresponds to an expansion of the region of the Bose-condensed phase for smaller $\alpha$.   
Furthermore, in the limit of $\alpha\to\infty$, where only the nearest-neighbor contribution of $\beta_{\alpha}$ survives, $\left(h/J \right)_{\text{c}}$ approaches $\pm 4$. This value is consistent with the result of previous works~\cite{vanOosten2001,Sachdev_1999}.

\subsection{Linear stability analysis for $\mathbf{K} \neq \mathbf{0}$}
\label{Excitation spectrum}

\begin{figure}
\includegraphics[width=1.0\linewidth]{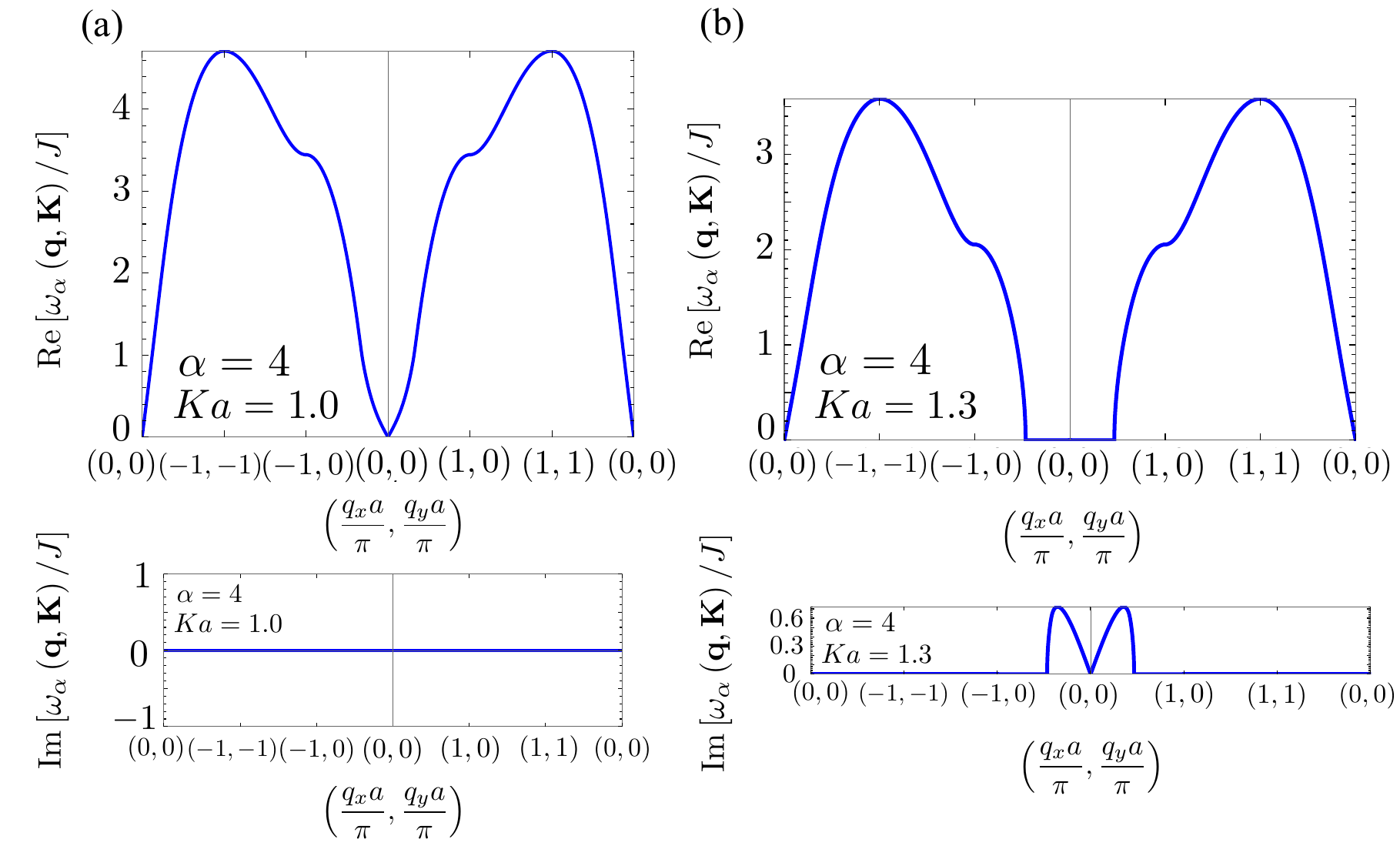}
\caption{Excitation spectra $\omega_\alpha(\mathbf{q},{\bf K})/J$ at $\alpha = 4$ and $n = 0.5$, where ${\bf K}={\bf e}_x K$. 
Panels (a) and (b) show the real and imaginary parts of the spectrum for $Ka = 1.0$ and $Ka = 1.3$, respectively. 
\label{excitationspectra}}
\end{figure}

In this section, we derive the excitation spectrum of current-carrying states in the Bose-condensed phase in order to analyze their stability. 
Specifically, we consider the following steady-state solution of Eqs.~(\ref{steady_equation_theta}) and (\ref{steady_equation_phi}), 
\begin{eqnarray}
\bar{\theta}_{j} = \theta_{0}, \,\, \bar{\phi}_{j} = -\mathbf{K}\cdot\mathbf{r}_{j},
\label{eq:currentSol}
\end{eqnarray}
where $\mathbf{K} = (K_{x}, K_{y})$ is the quasi-momentum of the BEC. In terms of the $S=1/2$ spin, it represents the gradient of the azimuthal angle 
such that the angle differences between nearest-neighbor spins along the $x$ and $y$ directions are $K_{x}a$ and $K_{y}a$, respectively. 

Through the mapping between the hard-core Bose-Hubbard model and the $S=1/2$ $XY$ model, 
the condensate wave function is expressed as
\begin{eqnarray}
\psi_j \equiv \langle \hat{b}_{j} \rangle = \langle \hat{S}^{-}_{j} \rangle
= \tfrac{1}{2}\sin\theta_{0}\,e^{-i\phi_{j}},
\label{eq:SForderparam}
\end{eqnarray}
which yields the condensate phase $\varphi_{j}=-\phi_{j}$ and the uniform condensate density $n^{\text{con}}_{j}=\tfrac{1}{4}\sin^{2}\theta_{0}=n ( 1 - n )$. 

From Eq.~(\ref{motion_theta}), one can easily derive the continuity equation for the HCB,
\begin{equation}
\frac{dn_j}{dt}
= 2\sum_{l\neq j} J_{j,l}(\alpha)\sqrt{n^{\rm con}_jn^{\rm con}_l} \sin\varphi_{jl},
\label{eq:continuity}
\end{equation}
where $\varphi_{jl}=\varphi_j-\varphi_l$.
The right-hand side of Eq.~(\ref{eq:continuity}) denotes the total particle current $I_j^{\rm tot}$ flowing into site $j$, meaning that the particle current carried by the BEC from site $l$ to $j$ 
is given by
\begin{eqnarray}
I_{j,l} = 2J_{j,l}(\alpha)\sqrt{n^{\rm con}_j n^{\rm con}_l}\sin\varphi_{jl}.
\end{eqnarray}
When the condensate density is homogeneous as in Eq.~(\ref{eq:SForderparam}), $I_{j,l}$ is written as
\begin{eqnarray}
I_{j,l}=-2n(1-n)J_{j,l}(\alpha)\sin\left(\mathbf{K}\cdot\left(\mathbf{r}_{l}-{\bf r}_j\right)\right)
\end{eqnarray}
Thus, the steady state of Eq.~(\ref{eq:currentSol}) with ${\bf K}\neq {\bf 0}$ indeed carries a current.

Substituting Eq.~(\ref{eq:currentSol}) into Eqs.~(\ref{MFenergy}) and (\ref{steady_equation_phi}), we obtain the energy per particle and the chemical potential as functions of ${\bf K}$ as
\begin{equation}
 \begin{split}
    \epsilon_{\alpha}({\bf K})
    & = \frac{ \mathcal{H}_{\text{MF}} + h\sum_{j}\langle \hat{S}_{j}^{z}\rangle }{ N }
    \\ & = -(1-n)J\,\gamma_{\alpha}({\bf K}),
    \label{energyband}
 \end{split}
\end{equation}
and
\begin{eqnarray}
h_\alpha({\bf K}) = (2n-1)J\gamma_\alpha({\bf K})
\end{eqnarray}
where $N$ is the number of particles and
\begin{equation}
\gamma_{\alpha}(\mathbf{K}) = \sum_{l\neq 0} \frac{J_{0,l}}{J}e^{i\mathbf{K}\cdot \mathbf{r}_l}.
\end{equation}
From Eq.~(\ref{energyband}), one can obtain the group velocity of the particle current,
\begin{eqnarray} 
{\bf v}_\alpha({\bf K}) = \nabla_{\bf K} \epsilon_{\alpha}({\bf K}).
\label{group_velocity}
\end{eqnarray}
where $\nabla_{\bf K} = {\bf e}_x \frac{\partial}{\partial K_x}+ {\bf e}_y\frac{\partial}{\partial K_y}$ and ${\bf e}_x$ (${\bf e}_y$) is the unit vector in the $x$-direction ($y$-direction). The group velocity is related to the current of the BEC through
\begin{eqnarray}
{\bf v}_\alpha({\bf K})=-\frac{1}{2n}\sum_{l\neq0}{\bf r}_l I_{0,l}.
\end{eqnarray}

Let us next calculate the excitation frequency $\omega$. Substituting Eq.~(\ref{eq:currentSol}) into Eqs.~(\ref{fluctuation_equation_theta}) and (\ref{fluctuation_equation_phi}) and anticipating the plane-wave solution for the fluctuations, 
\begin{eqnarray}  
\delta \theta_{j}(t) = \delta \theta_{\bf q}\,e^{ i \left( \mathbf{q} \cdot \mathbf{r}_{j}-\omega t \right) }, 
\delta \phi_{j}(t) = \delta\phi_{\bf q} \,e^{ i \left( \mathbf{q} \cdot \mathbf{r}_{j}-\omega t \right) },
\end{eqnarray}
one can solve Eqs.~(\ref{fluctuation_equation_theta}) and (\ref{fluctuation_equation_phi}) to obtain
\begin{equation}
 \begin{split}
\frac{\omega_\alpha(\mathbf{q},{\bf K})}{ J}  & =  \left( n - \frac{1}{2} \right) \bigl(  \gamma_{\alpha}( \mathbf{q} + \mathbf{K} ) - \gamma_{\alpha}( \mathbf{q} - \mathbf{K} )  \bigr)
\\ & + ~  \Biggl[  \biggl\{ \gamma_{\alpha}( \mathbf{K} ) - \frac{1}{2} \bigl( \gamma_{\alpha}( \mathbf{q} + \mathbf{K} ) +\gamma_{\alpha}( \mathbf{q} - \mathbf{K} ) \bigr) \biggr\}
\\ & \times \biggl\{ \gamma_{\alpha }( \mathbf{K} ) - \frac{1}{2} (  2 n - 1 )^{2} \bigl( ~ \gamma_{\alpha}( \mathbf{q} + \mathbf{K} )
\\ & + \gamma_{\alpha}( \mathbf{q} - \mathbf{K} ) \bigr) ~ \biggr\} \Biggr] ^{ \frac{1}{2} },
 \end{split}
\label{Excitationspectra}
\end{equation}
where ${\bf q}$ is the quasi-momentum of the excitation.
To avoid confusion, we emphasize that $\epsilon_{\alpha}(\mathbf{K})$ refers to the single-particle dispersion relation of the current-carrying condensate state with quasi-momentum $\mathbf{K}$, whereas $\omega_{\alpha}(\mathbf{q},\mathbf{K})$ denotes the excitation spectrum of a quasiparticle with quasi-momentum $\mathbf{q}$ in the condensate state with quasi-momentum $\mathbf{K}$.
Notice that when $\mathbf{K} = \mathbf{0}$, Eq.~(\ref{Excitationspectra}) reproduces the result of Refs.~\cite{Peter2012,Diessel2023}.

To be specific, we henceforth assume that the current is flowing in the $x$-direction, i.e., ${\bf K} = {\bf e}_xK$.
In Fig.~\ref{excitationspectra}, we show the real and imaginary parts of $\omega_\alpha(\mathbf{q},{\bf K})/J$ as a function of ${\bf q}$ at $\alpha=4$ and $n=0.5$ for $Ka=1.0$ and $1.3$.
At $Ka=1.0$, $\mathrm{Re}\left[ \omega_{\alpha}(\mathbf{q},{\bf K})/J \right] > 0$ and $\mathrm{Im}\left[ \omega_\alpha(\mathbf{q},{\bf K})/J \right] = 0$ (Fig.~\ref{excitationspectra} (a)), implying that the steady state is stable. In contrast, at $Ka=1.3$ (Fig.~\ref{excitationspectra} (b)), $\mathrm{Im}\left[ \omega_\alpha(\mathbf{q},{\bf K})/J \right] \neq 0$, implying that the steady state is dynamically unstable. The critical quasi-momentum $K_{\rm c}a$ of the DI is determined as the smallest positive value of $Ka$ at which $\mathrm{Im}\left[ \omega_\alpha(\mathbf{q},{\bf K})/J \right]$ first becomes nonzero.
At $n=0.5$, LI does not occur since $\mathrm{Re}[\omega_\alpha(\mathbf{q},{\bf K})/J]$ in Eq.~(\ref{Excitationspectra}) is always positive. 
For $n\neq0.5$, the critical quasi-momentum of LI is determined by finding the value of $Ka$ at which a region in $\mathbf{q}a$ space satisfying $\mathrm{Re}[\omega_\alpha(\mathbf{q},{\bf K})/J] < 0$ first appears.

\begin{figure}
\includegraphics[width=1.0\linewidth]{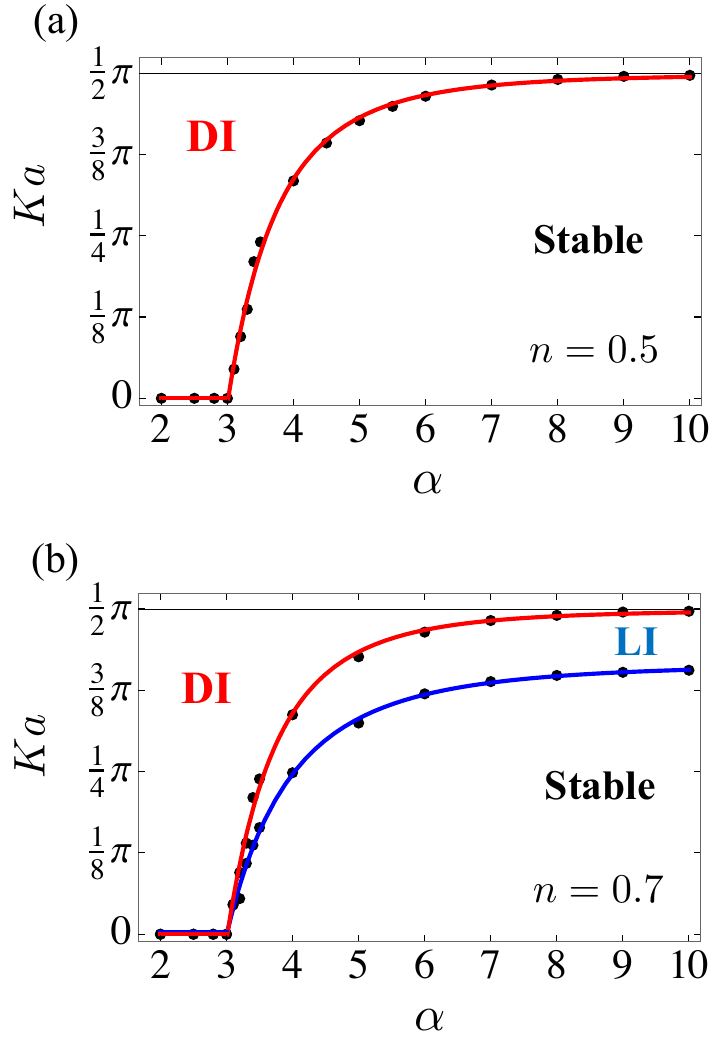}
\caption{
Stability phase diagram of the Bose-condensed state in the $(\alpha,Ka)$-plane.
Panels (a) and (b) correspond to $n = 0.5$ and $0.7$.  
Data points with red and blue interpolating curves represent the critical quasi-momenta at which the dynamical instability (DI) and Landau instability (LI) set in, respectively. 
\label{Critical_quai_momentum}}
\end{figure}

The stability phase diagrams in the $(\alpha,Ka)$-plane for $n=0.5$ and $n=0.7$ are shown in Figs.~\ref{Critical_quai_momentum}(a) and (b). 
In both cases, the fitting curve for the DI critical quasi-momentum converges toward $Ka = \pi/2$ in the limit $\alpha \to \infty$, corresponding to the nearest-neighbor hopping case~\cite{Smerzi2002,Wu_2003,Polkovnikov2005,Mun2007,Danshita2010}.
As $\alpha$ decreases, i.e., long-range hopping becomes more pronounced, both DI and LI critical quasi-momenta shift to smaller values. 
This demonstrates that stronger long-range hopping suppresses superfluidity, in the sense that the range of quasi-momenta over which the BEC can sustain stable flow becomes smaller. 
In particular, at $\alpha = 3$ corresponding to the case of Rydberg-atom arrays~\cite{Sylvain2019,chen2023continuous}, both DI and LI critical quasi-momenta vanish, implying that the Bose-condensed state cannot accommodate stable supercurrents in the region of $\alpha\leq 3$.

\begin{figure}
\includegraphics[width=1.0\linewidth]{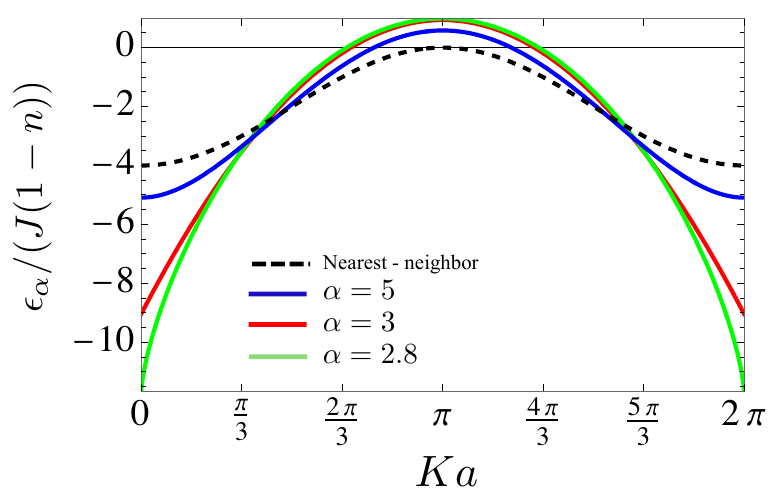}
\caption{
Single-particle energy band $\epsilon_{\alpha}$ defined in Eq.~(\ref{energyband}) for different decay exponents $\alpha$.
The horizontal axis represents quasi-momentum $Ka$ of the BEC.
The dashed black, blue, red, and green lines correspond to the nearest-neighbor hopping case
($\alpha \to \infty$), $\alpha = 5$, $\alpha = 3$ and $\alpha = 2.8$, respectively.
\label{Energyband}}
\end{figure}

\begin{figure}
\includegraphics[width=1.0\linewidth]{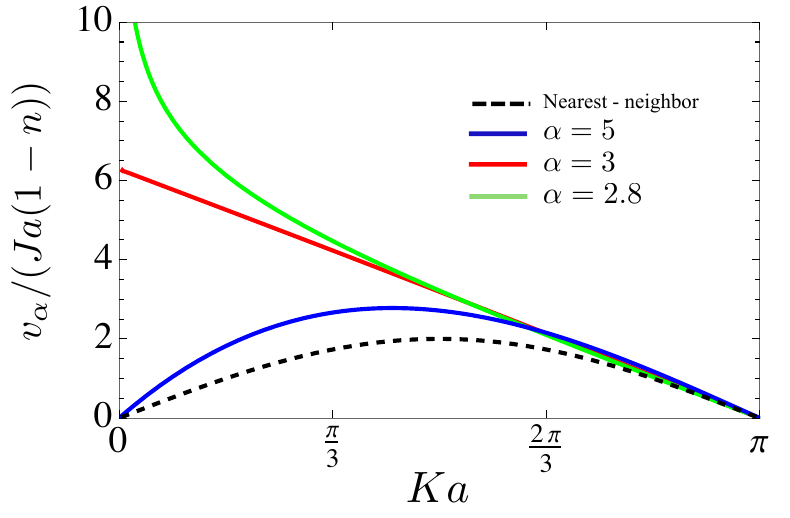}
\caption{
Group velocity $v_{\alpha}$ defined in Eq.~(\ref{group_velocity}) as a function of $Ka$ for different decay exponents $\alpha$.
The dashed black, blue, red, and green lines correspond to the nearest-neighbor hopping case ($\alpha \to \infty$), $\alpha = 5$, $\alpha = 3$, and $\alpha = 2.8$, respectively.
At $\alpha = 2.8$, $v_{\alpha}$ diverges in the limit $K\to 0$.
\label{Groupvelocity}}
\end{figure}

\begin{figure}
\includegraphics[width=1.0\linewidth]{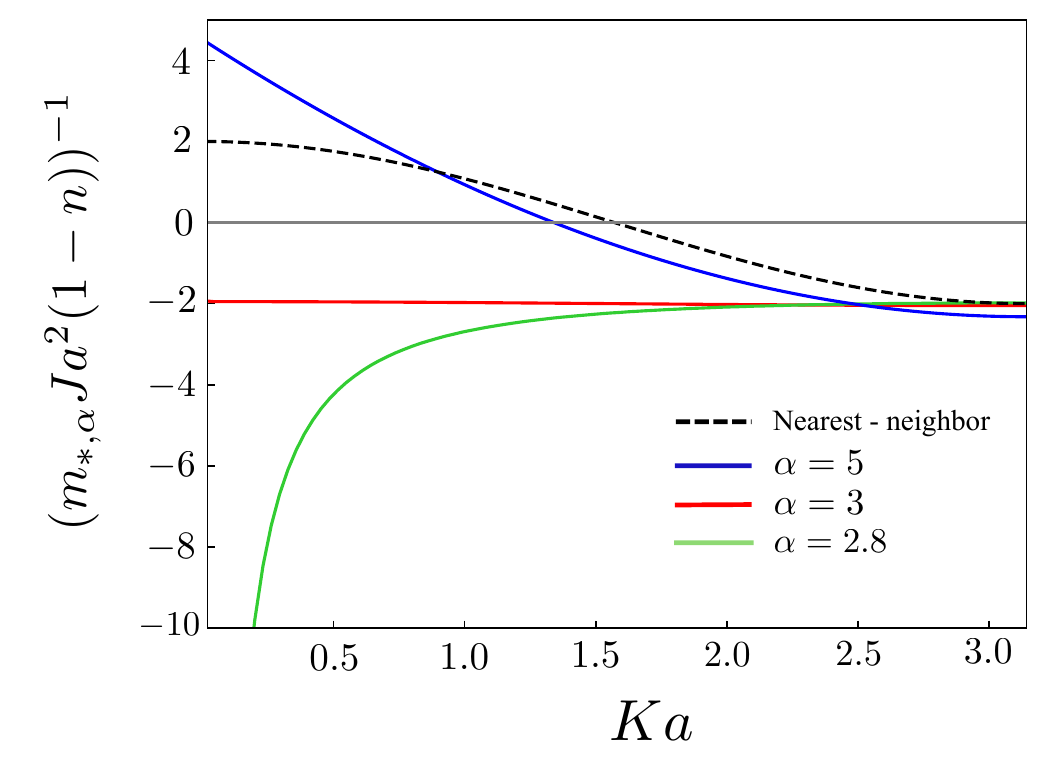}
\caption{
Effective mass $m_{\ast,\alpha}$ defined in Eq.~(\ref{eq:meff}) as a function of $Ka$ for different decay exponents $\alpha$.
The dashed black, blue, red, and green lines correspond to the nearest-neighbor hopping case ($\alpha \to \infty$), $\alpha = 5$, $\alpha = 3$, and $\alpha = 2.8$, respectively.
For $\alpha = 3$ and $\alpha = 2.8$, $m_{\ast, \alpha}$ is negative over the entire range of $Ka$, reflecting the negative curvature of $\epsilon_{\alpha}$.
At $\alpha = 2.8$, $m_{\ast, \alpha}$ diverges in the limit $K\to 0$.
\label{Effectivemass}}
\end{figure}

To elucidate how long-range hopping suppresses superfluidity and ultimately leads to its complete breakdown at $\alpha = 3$, we show in Fig.~\ref{Energyband} the single-particle energy band $\epsilon_{\alpha}(K)$ defined in Eq.~(\ref{energyband}).
The dashed black, blue, red, and green lines represent the nearest-neighbor hopping case ($\alpha\to \infty$), $\alpha = 5$, $\alpha = 3$, and $\alpha = 2.8$, respectively.
We see that for smaller $\alpha$, $\epsilon_{\alpha}(K)$ increases more steeply with $Ka$ and exhibits a more convex shape.
This behavior of $\epsilon_{\alpha}(K)$ can be understood intuitively in terms of the equivalent ferromagnetic $XY$ model with long-range interactions.
From this viewpoint, stronger long-range ferromagnetic interactions (i.e., smaller $\alpha$) increase the energy cost required to twist the azimuthal angle of neighboring spins by $Ka$.
In the present system, this increased energy cost manifests itself as a steeper and more convex rise of $\epsilon_{\alpha}(K)$ with $Ka$ compared to the nearest-neighbor or larger-$\alpha$ cases, resulting in a shrinking concave region that eventually becomes fully convex at $\alpha = 3$.
Consequently, the inflection point of $\epsilon_{\alpha}(K)$ shifts to smaller values of $Ka$ as $\alpha$ decreases. 
The convexity of $\epsilon_{\alpha}(K)$ leads to DI in the following manner. The effective mass $m_{\ast,\alpha}$ of the particle on the lattice is obtained from the inverse of second derivative of the energy band $\epsilon_{\alpha}(K)$ with respect to $K$, i.e.,
\begin{eqnarray}
m_{\ast, \alpha} = \left(\frac{\partial^2 \epsilon_\alpha}{\partial K^2}\right)^{-1}
\label{eq:meff}
\end{eqnarray}
and the sound velocity $c$ is expressed as 
\begin{eqnarray}
c = \frac{1}{\sqrt{\kappa m_{\ast,\alpha}}},
\label{eq:sound}
\end{eqnarray}
where $\kappa$ is the compressibility~\cite{Kramer2003}.
In order to obtain the effective mass $m_{\ast,\alpha}(K)$, we calculate the group velocity $v_\alpha(K)=\frac{\partial \epsilon_\alpha}{\partial K}$ in the $x$ direction, which is plotted in Fig.~\ref{Groupvelocity}. Using the relation $m_{\ast,\alpha}^{-1}=\frac{\partial v_\alpha}{\partial K}$, we further calculate $m_{\ast,\alpha}(K)$ and show those for different values of $\alpha$ in Fig.~\ref{Effectivemass}.
In the nearest-neighbor and $\alpha=5$ cases, $m_{\ast,\alpha}^{-1}$ changes from positive to negative at a finite value of $Ka$, indicating that $\epsilon_{\alpha}(K)$ has an inflection point there.
This value of $Ka$ coincides with the critical quasi-momentum $K_{\rm c}a$ for the DI shown in Fig.~\ref{Critical_quai_momentum}.
For $\alpha=3$ and $\alpha=2.8$, $m_{\ast,\alpha}$ is negative for all $Ka$, reflecting the negative curvature of the single-particle dispersion $\epsilon_{\alpha}(K)$.
Eq.~(\ref{eq:sound}) means that the negative effective mass leads to an imaginary sound velocity.
Since we numerically confirm that DI of the current-carrying state is caused by normal modes with long wavelength, the inflection point of $\epsilon_{\alpha}(K)$ corresponds to the critical quasi-momentum $K_{\rm c}a$ at which the DI sets in.
In short, the long-range nature of the hopping enhances the convexity of  $\epsilon_{\alpha}(K)$ such that it suppresses the superfluid critical quasi-momentum for DI with decreasing $\alpha$, eventually making $K_{\rm c}a$ vanish at $\alpha = 3$.

It is worth noting that, when $\alpha = 3$, the group velocity $v_{\alpha}(K)$ in the $x$ direction  remains finite in the limit $K\to 0$, and $\lim_{K\to 0} v_\alpha$ diverges for $\alpha < 3$, as shown in Fig.~\ref{Groupvelocity}.
In this sense, the fact that the critical quasi-momentum for $\alpha \leq 3$ is zero does not mean that the corresponding critical velocities are zero as well.
Instead, the critical value of the group velocity is ill defined in the limit $K\to 0$ for $\alpha < 3$, while the quasi-momentum $K$, which labels the phase-twisted stationary state, remains a well-defined control parameter.
Therefore, the onset of instability for $\alpha \leq 3$ should be identified experimentally from the decay of the prepared finite-$K$ state or the growth of long-wavelength modulations, rather than from a transport velocity.
In Rydberg-atom arrays and trapped-ion systems, the $XY$ ferromagnetic ordered state with finite $K$ can be prepared by first generating an $XY$ ferromagnetic ordered state with $K=0$ via an adiabatic preparation, as was done in Ref.~\cite{chen2023continuous}, and then imprinting a linearly modulated phase $\varphi_j = {\bf K}\cdot {\bf r}_{j}$ of the condensate, where ${\bf K} = {\bf e}_xK$, via site-dependent AC Stark shifts~\cite{Bornet2023,PhysRevLett.132.263601,PhysRevLett.119.053202,Schindler_2013,PhysRevA.94.042308,ChWunderlich_2003} that act as single-particle potentials for hard-core bosons.

\subsection{\label{Results and discussion} Critical behavior of the critical quasi-momentum}

\begin{figure}
\includegraphics[width=1.0\linewidth]{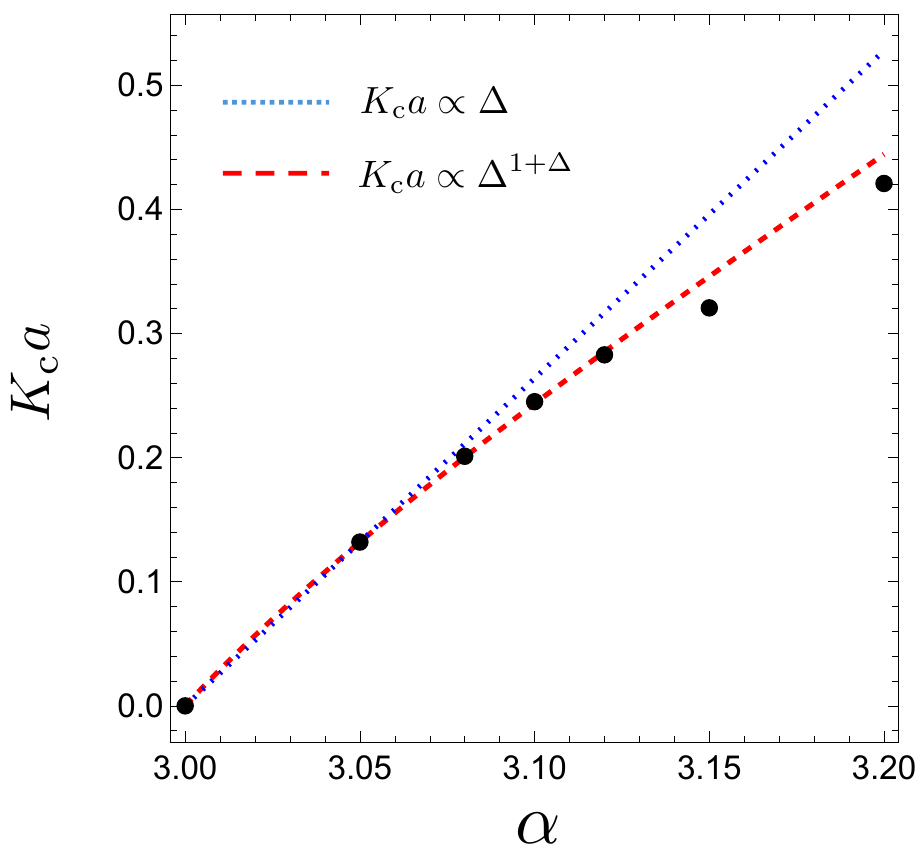}
\caption{
Comparison between the analytical prediction and numerical results for the DI critical quasi-momentum $K_{\mathrm{c}}a$ near $\alpha = 3$. 
Symbols denote numerical results obtained from the excitation spectrum of Eq.~(\ref{Excitationspectra}). 
The red dashed line represents the analytical scaling $K_{\rm c}a \propto \Delta^{1 + \Delta}$ with $\Delta = \alpha - 3$, while the blue dotted line shows the linear scaling $K_{\rm c}a \propto \Delta$ for reference. 
The proportionality constants in both analytical curves are fixed by the data point at $\alpha = 3.05$. 
\label{Comp_fig}}
\end{figure}

At $\alpha = 3$, the DI critical quasi-momentum $K_{\rm c}a$ becomes zero, and its critical behavior in the vicinity of this point can be derived analytically as follows.
As mentioned in the previous section, the DI critical quasi-momentum $K_{\rm c}a$ is determined by the condition $m_{\ast, \alpha}^{-1} = 0$.
In the limit of $\alpha \to \infty$, this condition yields $K_{\rm c}a = \pi/2$, consistent with the nearest-neighbor hopping case~\cite{Smerzi2002,Wu_2003,Polkovnikov2005,Mun2007,Danshita2010}. 
For small quasi-momenta $K \ll 1/a$, the discreteness of the lattice becomes negligible so that $\gamma_{\alpha}(\mathbf{K})$ can be approximated by its continuum limit, which can be written in the integral form
\begin{eqnarray}
    \gamma_{\alpha}(\mathbf{K}) &\simeq &
    \frac{1}{a^{2}} \int_{a}^{\infty} d\mathbf{r} \left( \frac{a}{r} \right)^{\alpha} e^{i\frac{\mathbf{r}}{a}\cdot\mathbf{K}a}
    \nonumber \\
    & =& \int_{1}^{\infty} d\rho\, \frac{1}{\rho^{\alpha-1}} J_{0}(\eta\rho)
   \nonumber \\
   &&  \!\!\!\!\!\!\!\!\!\!\!\!\!\!\!\!\!\!\!\!\!\!\simeq 2\pi\!\left(-\frac{1}{\alpha-2} - \eta^{\alpha-2}A + \frac{\eta^{2}}{4(4-\alpha)} + O(\eta^{4})\right),
\end{eqnarray}
where $\rho = r/a$, $\eta = Ka$, and $A = \frac{\alpha\,\Gamma(-\frac{\alpha}{2})}{2^{\alpha}\Gamma(\frac{\alpha}{2})}$.
Here, $J_{0}$ denotes the Bessel function of the first kind.
Assuming that $K_{\rm c}a \ll 1$ in the vicinity of $\alpha = 3$, the condition $m_\ast^{-1}=0$ gives
\begin{equation}
    K_{\rm c}a \propto \Delta^{1+\Delta},
    \quad \text{with} \quad \Delta = \alpha - 3.
\end{equation}

In Fig.~\ref{Comp_fig}, we compare the analytical prediction $K_{\rm c}a \propto \Delta^{1+\Delta}$ with numerical data (symbols) to examine their agreement near $\alpha = 3$. 
For reference, we also plot the linear scaling $K_{\rm c}a \propto \Delta$ (blue dotted line). 
The proportionality constants in both analytical expressions are fixed so that they reproduce the numerical value of $K_{\rm c}a$ at $\alpha = 3.05$. 
The linear scaling begins to deviate from the numerical data around $\alpha \simeq 3.10$, 
whereas the analytical result $K_{\rm c}a \propto \Delta^{1+\Delta}$ remains consistent with the numerical data.
This confirms that the scaling $K_{\rm c}a \propto \Delta^{1+\Delta}$ reproduces the trend of the numerical data more accurately. 
We emphasize that the critical exponent explicitly depends on $\Delta$, in contrast to short-range models where the exponent is typically constant.
This nontrivial behavior originates from the long-range nature of the hopping.

We expect our mean-field conclusion that $\alpha=3$ is the critical value for stability to remain qualitatively valid beyond the mean-field regime.
This expectation is supported by recent large-scale quantum Monte Carlo results for related two-dimensional long-range Heisenberg models with interactions decaying as $1/r^{\alpha}$~\cite{PhysRevResearch.5.033046}, which showed that even in the presence of beyond-mean-field fluctuations, the low-$q$ sector of the excitation spectrum remains in good agreement with analytical predictions based on spin-wave theory~\cite{Diessel2023} over a wide range of $\alpha$.
Although those studies were performed for Heisenberg models rather than for $XY$ models, they still provide a useful benchmark for the present system.
Since the present instability is governed by such a low-$q$ sector of excitations, qualitative properties of the corresponding critical quasi-momentum $K_{\mathrm c} a$, including the critical behavior near $\alpha=3$, are also expected to be robust, although a quantitative shift of the $K_{\mathrm c} a$ values beyond mean field cannot be excluded.
The beyond-mean-field shift in the critical quasi-momentum may be captured by a cluster mean-field theory as was done for hard-core bosons with long-range density-density interactions in Ref.~\cite{Yamamoto_2011} and such an analysis is left for future work.

\section{\label{Summary} SUMMARY}

Using a mean-field theory, we have studied the hard-core Bose-Hubbard model on a square lattice with long-range hopping decaying algebraically with the distance $r$ as $\propto r^{-\alpha}$ in order to analyze the stability of current-carrying states in the Bose-condensed phase.
Taking advantages of the equivalence between the hard-core Bose-Hubbard model and the spin-1/2 $XY$ model, we calculated the excitation spectrum and determined the critical quasi-momenta for both LI and DI.
We have shown that when $\alpha$ decreases from a large value, the critical quasi-momemta monotonically decreases and eventually vanish at $\alpha=3$, which corresponds to the case of Rydberg-atom arrays~\cite{Sylvain2019,chen2023continuous,Chen2025}. 
We have attributed this behavior to the increase in the energy cost of the phase twisted configuration in the current-carrying states.
Furthermore, we found that the DI critical quasi-momentum $K_{\rm c}a$ exhibits the scaling behavior $K_{\rm c}a \propto \Delta^{1+\Delta}$ near $\alpha=3$, with $\Delta=\alpha-3$. 
The emergence of the unconventional critical behavior, in which the exponent depends explicitly on $\Delta$, reflects the long-range nature of the hopping. 

While we theoretically explored a large region of $\alpha$, a system with $\alpha > 3$ has not been experimentally realized except for $\alpha=6$ and $\alpha\rightarrow \infty$. It will be interesting to seek experimental ways of how the regime $\alpha > 3$ can be realized with current experimental techniques.
Moreover, while we focused on the superfluid critical quasi-momentum as a signature of the superfluidity, the nature of the superfluidity emerges also as a finite superfluid fraction and the quantization of vortices~\cite{Pitaevskii}. It will be meaningful to investigate these properties in the system with the long-range hopping addressed in the present work. In this direction, a recent theoretical study has addressed the superfluid fraction of the soft-core Bose-Hubbard model with the long-range hopping in one dimension~\cite{Gupta2025}. 

\begin{acknowledgments}
We thank Masaya Kunimi and Takafumi Tomita for useful discussions.
This work is financially supported by MEXT Q-LEAP  [Grant No.~JPMXS0118069021], 
JST FOREST [Grant No.~JPMJFR202T], 
JST ASPIRE [Grant No.~JPMJAP24C2], JSPS KAKENHI [Grant No.~JP26K00639], and JST SPRING [Grant No.~JPMJSP2139].

\end{acknowledgments}

\section*{DATA AVAILABILITY}

The data that support the findings of this article are not publicly available. The data are available from the authors upon reasonable request.

\bibliography{rydbergrefs}

\end{document}